\documentclass[aip, rsi, reprint, graphicx]{revtex4-1}
\usepackage{amsmath}
\usepackage{siunitx}
	\DeclareSIUnit{\prhz}{\ensuremath{\per\sqrt{\text{\hertz}}}}
	\DeclareSIUnit\rad{\radian}
\usepackage{graphicx}
\graphicspath{ {figures/} }
\usepackage[breaklinks = true, colorlinks = true]{hyperref}
\usepackage[capitalize]{cleveref}

\draft 

\newcommand{\pendBestIfoN}{\SI{200}{\femto\N\prhz}}

\newcommand{\pendBestIfoA}{\SI{4e-13}{\m\per\square\s\prhz}}
\newcommand{\pendBestIfoF}{\SI{2}{\milli\Hz}}

\newcommand{\noiseRatio}{20}

\newcommand{\pendBestCapA}{\SI{2e-12}{\m\per\square\s\prhz}}
\newcommand{\pendBestACapF}{\SI{0.5}{\milli\Hz}}

\newcommand{\sensCap}{\SI{30}{\nm\prhz}}
\newcommand{\sensCapPhi}{\SI{100}{\nano\rad\prhz}}
\newcommand{\sensIfo}{\SI{0.5}{\nm\prhz}}
\newcommand{\sensIfoPhi}{\SI{1}{\nano\rad\prhz}}

\begin{document}


\title{A New Torsion Pendulum for Gravitational Reference Sensor Technology Development} 



\author{Giacomo Ciani}
\email{ciani@phys.ufl.edu}
\author{Andrew Chilton}
\author{Stephen Apple}
\author{Taiwo Olatunde}
\author{Michael Aitken}
\author{Guido Mueller}
\author{John W. Conklin}
\affiliation{University of Florida, Gainesville, FL-32611, USA}



\begin{abstract}
We report on the design and sensitivity of a new torsion pendulum for measuring the performance of ultra-precise inertial sensors and for the development of associated technologies for space-based gravitational wave observatories and geodesy missions.
The apparatus comprises a \SI{1}{\m}-long, \SI{50}{\um}-diameter, tungsten fiber that supports an inertial member inside a vacuum system.
The inertial member is an aluminum crossbar with four hollow cubic test masses at each end.
This structure converts the rotation of the torsion pendulum into translation of the test masses.
Two test masses are enclosed in capacitive sensors which provide readout and actuation.
These test masses are electrically insulated from the rest of the cross-bar and their electrical charge is controlled by photoemission using fiber-coupled ultraviolet light emitting diodes.
The capacitive readout measures the test mass displacement with a broadband sensitivity of \sensCap{}, and is complemented by a laser interferometer with a sensitivity of about \sensIfo{}.
The performance of the pendulum, as determined by the measured residual torque noise and expressed in terms of equivalent force acting on a single test mass, is roughly \pendBestIfoN{} around \pendBestIfoF{}, which is about a factor of \noiseRatio{} above the thermal noise limit of the fiber.
\end{abstract}


\maketitle 


\section{Introduction} \label{sec:intro}
Precision inertial sensors for spacecraft with acceleration noise performance below the \si{\nm\per\square\s\prhz} level at frequencies below \SI{1}{\Hz} are used for Earth geodesy missions \cite{CHAMP, GRACE, GOCE}, space-based gravitational wave observatories \cite{LISA, gravUniverse}, fundamental physics missions \cite{GPB, GPBattControl, microscope}, and for precision navigation and orbit determination \cite{nguyen2015, leitner2003}.

A gravitational reference sensor (GRS) is a class of inertial sensor that provides the primary measurement for drag-free control systems \cite{lange1964, debra2011}.
It typically consists of a free-floating dense metallic test mass (TM) surrounded by a housing that protects the test mass from external forces generated either by the space environment or by the spacecraft itself.
Sensors are used to measure the position of the TM with respect to the housing which is fixed to the spacecraft.
Thrusters on board the spacecraft are then commanded to keep the spacecraft centered with respect to the TM.
Thus the spacecraft and test mass will follow a near-perfect gravitational orbit in space.
By measuring the spacecraft's orbit or the difference between two drag-free spacecraft orbits, we learn about the spatial and temporal variations in the gravitational field or field gradient.

The performance of a gravitational reference sensor is determined by the level of residual TM acceleration measured in units of \si{\m\per\square\s\prhz} over a relevant frequency band.
A heavier test mass, larger gaps between the TM and its housing (which helps reduce unwanted surface interactions), and a quieter thermal, gravitational and electromagnetic spacecraft environment all lead to lower acceleration noise levels. However, larger gaps and heavier TM also reduce the sensitivity of capacitive sensors as well as the ability to actuate the TM. 

Testing sub-\si{\nm\per\square\s\prhz} inertial instruments at frequencies below \SI{1}{\Hz} is inherently challenging in the 1~\textit{g} laboratory environment.
However, laboratory torsion pendulums have been successfully used to test several aspects of precision inertial sensors.
The University of Trento has been operating such pendulums for a number of years \cite{TrentoSingleTM,TrentoFourTM}, studying the specific geometry of the GRS\cite{LPF_GRS} for the Laser Interferometer Space Antenna (LISA)\cite{LISAprop2017,gravUniverse} mission and its technology demonstration precursor mission LISA Pathfinder (LPF)\cite{LPF}.
More recently, a double torsion pendulum also aimed at testing realistic sensor geometries was developed at the University of Naples; it features one sensitive translational and one sensitive rotational degree-of-freedom of the TM for the study of cross-coupling effects \cite{PETER}.
Similar pendulums, although using a sensor geometry less representative of a flight-like GRS, have been also built at the Huazhong University of Science and Technology in Wuhan, China \cite{WuhanSingle,WuhanDouble}.
Torsion pendulums developed at the University of Washington are used to test the underlying physics governing the performance of precision inertial sensors \cite{UW}.

Here, we report on the design of a unique torsion pendulum developed at the University of Florida to specifically test new technologies for the LISA GRS. However, this research is more broadly applicable to precision space inertial sensors in general. Our design largely builds upon the experience and configuration of the facility at the University of Trento \cite{GiacThesis}, although with significant differences.

LISA will be the first space-based gravitational wave observatory, consisting of three Sun-orbiting drag-free spacecraft forming a roughly equilateral triangle with an arm length of 2.5\,\si{\giga\m}.
Each spacecraft will house two gravitational reference sensors similar to the ones employed in LPF.
The internal free-floating test masses are \SI{46}{\mm}, \SI{1.9}{\kg} cubes fabricated from an Au-Pt alloy and coated in Au.
The gaps between the test mass and its Au-coated, molybdenum housing range from \SIrange{3}{4}{\mm}.
The frequency band of interest for LISA is about \SIrange{0.1}{100}{\milli\Hz}, which also defines the most relevant measurement band for our torsion pendulum.
The most stringent acceleration noise requirement for LISA in terms of amplitude spectral density is \SI{3}{\femto\m\per\square\s\prhz} at \SI{\sim 0.3}{\milli\Hz}. 

\section{Apparatus Description} \label{sec:mechdesign}
\begin{figure} 
	\includegraphics[width = \columnwidth]{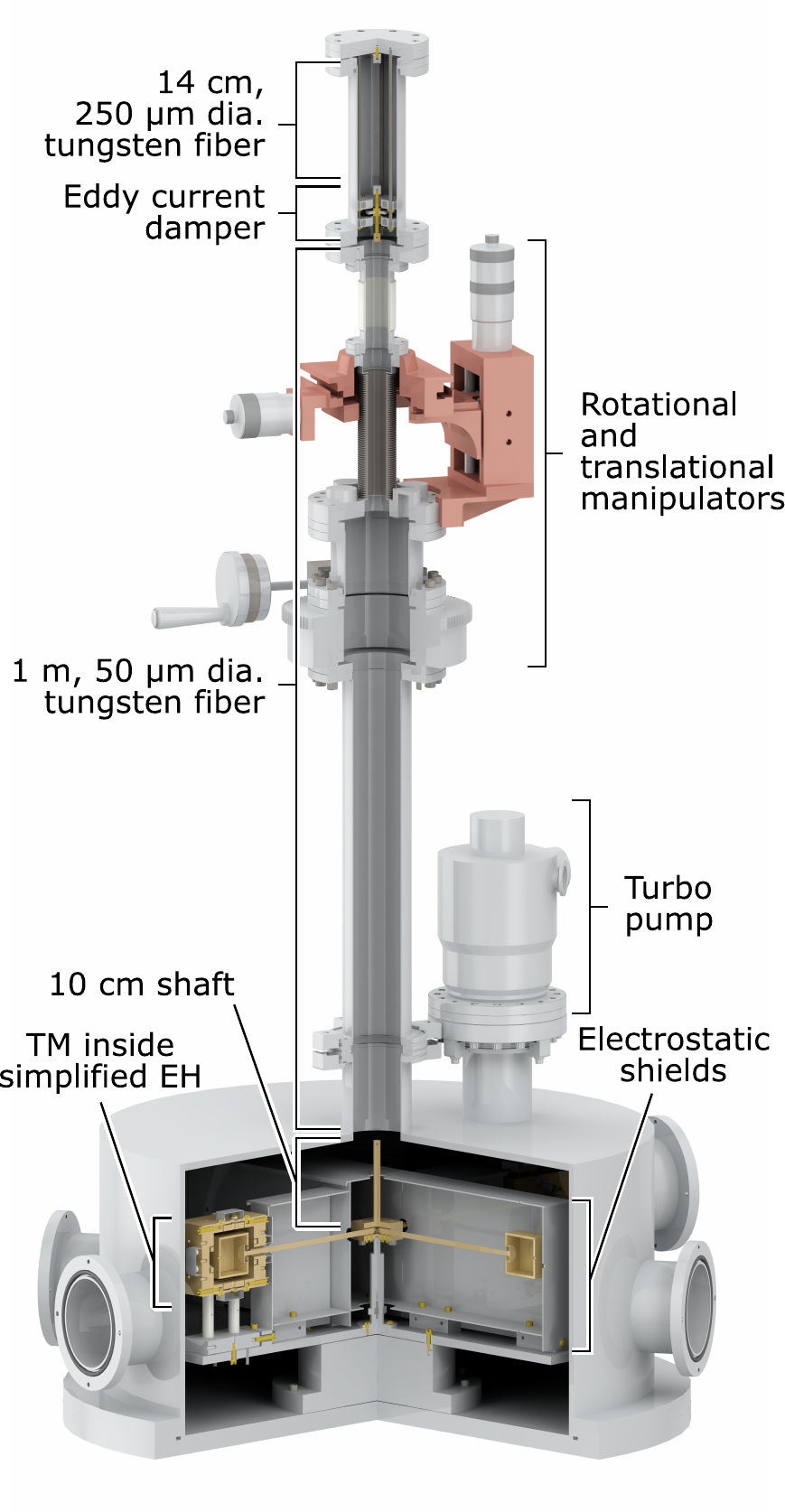}
	\caption{\label{fig:pendulumCAD}
		(Color online) Section view of a CAD model of the UF torsion pendulum. The inertial member and one of the simplified EH enclosing one hollow TM are visible in gold color a the bottom. The location and length of the upper and lower fibers, too thin to be visible, are indicated by the annotations.}
\end{figure}

The University of Florida (UF) torsion pendulum, depicted in \cref{fig:pendulumCAD}, consists of a rigid inertial member suspended by a thin torsional fiber.
The motion of the pendulum is measured by both capacitive and interferometric sensors.
The system is installed inside a vacuum chamber, which is equipped with various positioning stages to precisely align the pendulum to the sensors.
\subsection{Mechanical design}
\begin{figure}
	\includegraphics[width = \columnwidth]{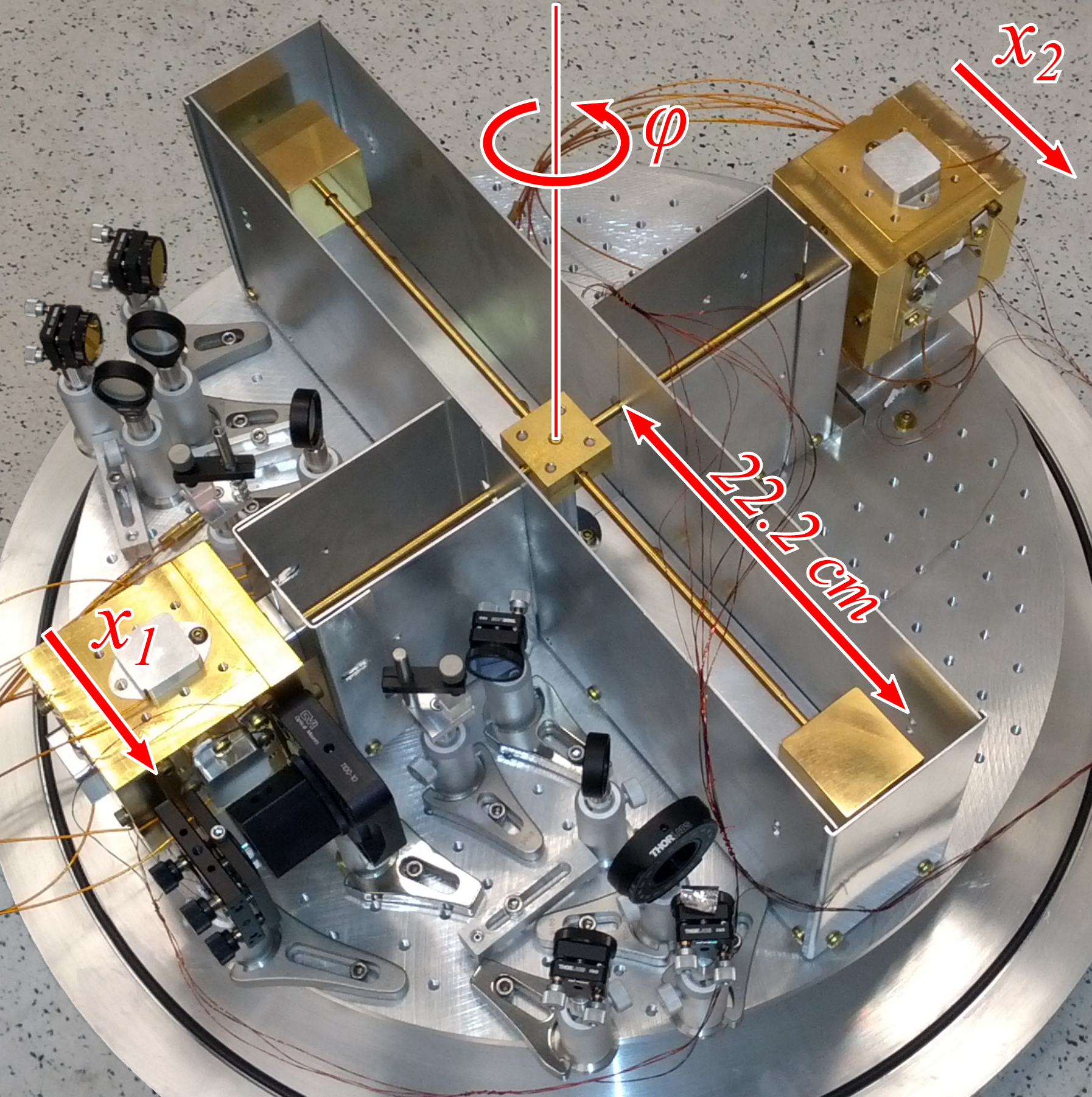}
	\caption{\label{fig:inertialMember}
		(Color online) A picture of the pendulum inertial member surrounded by some of the electrostatic shields. Two of the four TM are visible, while the other two are embedded in the simplified EHs (in the lower left and upper right corners of the picture). The optical components of the readout interferometer are also shown. The most relevant degrees of freedom are labeled in the picture. Also indicated is the arm-length, to provide an overall scale for the photograph.}
\end{figure}

The inertial member, shown in \cref{fig:inertialMember}, is comprised of a crossbar which supports four cubic TMs.
The TMs are light-weighted to reduce the mass of the inertial member and therefore the diameter and associated thermal noise of the torsion fiber.  
Each hollow \SI{46}{\mm} TM is fabricated from five individual aluminum parts bonded together with EPO-TEK H20E conductive epoxy using internal surfaces such that the epoxy is not visible from the outside.
The aluminum surfaces are gold-coated for electrostatic uniformity.
The distance between the center of each TM and the center of the crossbar is \SI{22.2}{\cm}, and the total mass of the inertial member is \SI{477}{\g}.
Two of the TMs are surrounded by electrode housings (EHs), which are simplified versions of the LISA GRS electrode housing.
These two TMs are connected to the main pendulum structure via quartz rings, which provide electrical insulation;
this is necessary to simulate the properties of the GRS, which rely on an electrically floating TM.

The inertial member is completely surrounded by conductive grounded shields, which prevent electrostatic interactions with charge that may accumulate on dielectric surfaces (cables, vacuum windows and other non-metallic parts of the apparatus).
\begin{table}
	\caption{Most relevant physical parameters of the torsion pendulum.}
	\label{tab:params}
	\begin{tabular}{ll c c}
		\hline\noalign{\smallskip}
		Parameter 						& Symbol 	& Value \\
		\hline\noalign{\smallskip}
		Total pendulum length			& $L_t$		& \SI{1.30}{m} \\
		Torsion fiber length 			& $L$ 		& \SI{1.00}{\m} \\
		Fiber shear modulus 			& $E_r$ 	& \SI{161}{\giga\Pa} \\
		Fiber Young's modulus			& $Y$		& \SI{400}{\giga\Pa} \\
		Rotational stiffness			& $\Gamma$ 	& $\sim$\SI{e-7}{\N\m\per\rad} \\
		Elongation stiffness			& $k$		& \SI{785}{\N\per\m} \\
		Mass 							& $m$ 		& \SI{477}{\g} \\
		Crossbar arm length 			& $a$ 		& \SI{22.2}{\cm} \\
		M. of Inertia (around $z$)		& $I_r$		& \SI{8.26e-3}{\kg\m\squared} \\
		M. of Inertia (around $x$ or $y$)	& $I_t$	& \SI{1.65e-2}{\kg\m\squared} \\
		Lower shaft length				& $h$		& \SI{10}{\cm} \\
		\noalign{\smallskip} \hline
	\end{tabular}
\end{table}

The suspension fiber is a \SI{1}{\m} long, 99.95\% purity tungsten fiber with a diameter of \SI{50}{\um}.
At each end, the fiber is epoxied to a custom fiber attachment cap provided with a standard thread.
The lower torsion fiber cap is threaded onto the top of a vertical shaft on the inertial member (not shown in \cref{fig:inertialMember}) which elevates the attachment point to a height $h = \SI{10}{\cm}$ above the center of mass of the pendulum.
A magnetic damper is used to damp out the simple pendulum oscillation or \textit{swing} mode.
At its upper end, the main fiber is attached to a horizontal aluminum disc, which is in turn hung from a shorter (\SI{14}{\cm}), thicker (\SI{250}{\um} diameter) fiber eventually fixed to the top of the vacuum chamber.
The electrically conducting disk is immersed in an inhomogeneous magnetic field generated by a set of neodymium magnets fixed to the vacuum chamber, just above and below the disk.
When swinging, the two fibers behave as a single, longer fiber, causing the disk to move in the magnetic field and generating eddy currents which dissipate energy due to the non-zero resistance of aluminum.
Since the torsional stiffness of a fiber is proportional to its radius to the fourth power, the much thicker upper fiber experiences negligible deformation during rotational oscillation and does not impact the dynamics or noise of the torsional degree-of-freedom.
The current swing mode decay time is slightly longer than one hour.
We have however recently developed and tested an improved magnet cage design, which we plan to install during an upcoming upgrade of the facility.
The new magnet cage is expected to allow us to obtain decay times as low as \SI{150}{\s} (adjustable), while also providing larger clearance for the disk, thus minimizing the risk of mechanical interference.
The most relevant parameters of the mechanical design are summarized in \cref{tab:params}. 

\begin{figure}
	\includegraphics[width = 0.5\columnwidth]{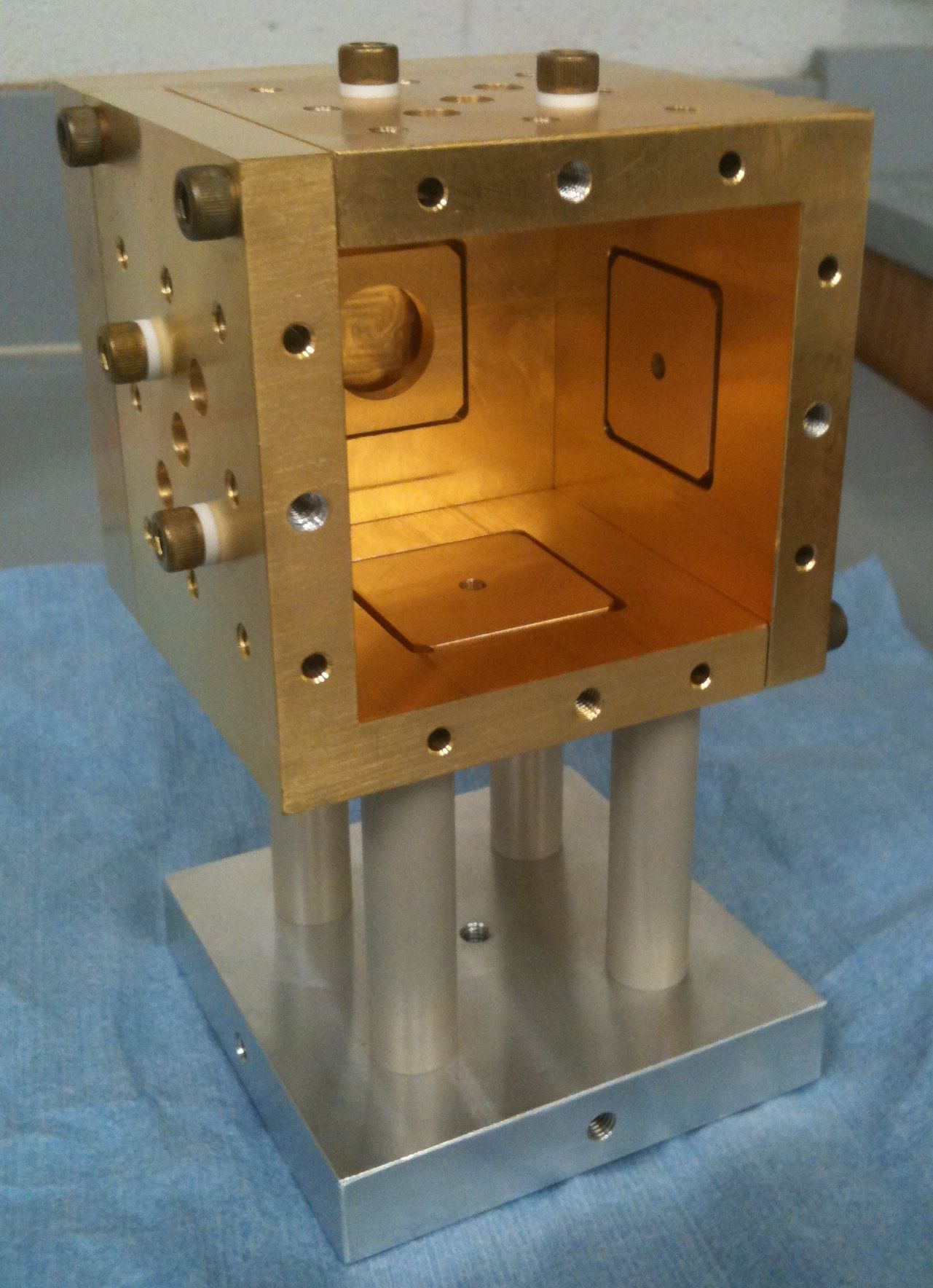}
	\caption{\label{fig:simpGRS}
		(Color online) One of two simplified EH integrated with the torsion pendulum, mounted on its supporting baseplate via four Peek legs for electrical and thermal insulation. The face towards the center of the pendulum has been removed to show the interior. Note the hole in the electrode on the far face, which for symmetry reasons mirrors the hole present in the missing face, needed for the shaft supporting the TM.}
\end{figure}

The apparatus employs two simplified EHs installed on opposite TMs to read out and control the pendulum position.
A simplified EH, depicted in \cref{fig:simpGRS}, includes a total of 6 electrodes centered on each of the interior faces.
This arrangement allows for measuring all three translational degrees of freedom of the TM, although only two of them simultaneously.
The electrodes are electrically isolated from the structure of the housing with ceramic bushings.
Like the TMs, the electrodes and the structure of the housing are made of gold-coated aluminum.
In the sensitive direction $x$ of each TM (corresponding to the pendulum rotation around the fiber's axis, see \cref{fig:inertialMember}), the gaps between the electrodes and test mass surface are \SI{8}{\mm}.
Since surface disturbances scale with the inverse of the gap size to some power, depending on the particular disturbance, this relatively large gap value reduces the simplified EH's contribution to the overall noise compared to more flight-like EH designs;
a realistic design will in fact likely have smaller gaps and can be tested by swapping it with one of the simplified EH units.

We designed the inertial member such that all of its internal resonant frequencies are well above our measurement band, and we can treat it as a rigid body.
We are mostly interested in the rotation $\phi$ around the axis defined by the torsional fiber, which is associated with the differential displacement $x_1-x_2$ of two opposing TMs with respect to their individual housings.
The displacements $x_1$ and $x_2$ are also sensitive to the swing mode in the $x$ direction, which is ideally common to both measurements.
The other degrees of freedom, swinging in the $y$ direction, changes in the height $z$ of the pendulum (bounce mode) and rotations around the $x$ and $y$ axis through the center of the crossbar do not in first order affect $x_1$ or $x_2$.

Ultraviolet (UV) light emitting diodes (LEDs) are coupled to optical fibers which end in dedicated holes in the electrode housing such that UV light can be used to control the charge of the test mass \cite{LISA_CMS}.
There are 3 available optical fiber ports located on each simplified EH: one illuminates the TM, one illuminates the electrode housing, and one illuminates both. This allows for the testing of different charge control strategies.

The two EHs, accompanied by precision positioning aids, the electrostatic shields and the interferometer hardware (described in \cref{sec:ifo}) are installed on top of a \SI{27}{\cm}-radius round aluminum platform, supported by a spacer designed to be replaced by a motorized rotational stage that will provide remote and precise centering capabilites in $\phi$.
The entire apparatus is located inside a \SI{60}{\cm}-diameter vacuum chamber, shown in \cref{fig:pendulumCAD}, which includes a tall tube that encloses the suspension fiber.
The top section of the tube is attached to both a rotational manipulator to allow adjustment of the equilibrium angle of the pendulum, and a three degree-of-freedom translational manipulator to allow for centering and calibration.
Above the rotational and translational manipulators, a ceramic section of the vacuum tube electrically isolates the top plate from the rest of the chamber, which is grounded through the vacuum pumps.
This allows the electrical potential of the torsion fiber and the pendulum inertial member to be arbitrarily controlled and partially isolated from the noisy electrical ground of the electrical equipment.
Currently the base pressure inside the vacuum chamber is about \SI{4e-6}{\milli\bar}.
Previous measurements on similar facilities have shown that Brownian noise associated with residual gas damping can be an important contribution to the measured noise \cite{gasDamp};
by rescaling those results to our pendulum geometry, accounting for different arm-length and sensors arrangement, we estimate the total torque noise from this source to be at the level of about \SI{3e-14}{\m\per\square\s\prhz}, exceeding the fiber thermal noise at frequencies of about \SI{1}{\milli\Hz} and above.
Although this indicates that we will eventually need to improve the base pressure, it is not currently limiting the performance of the apparatus.

\subsection{Environmental Monitoring and Isolation}

Several measures have been taken to monitor environmental effects and, in some cases, reduce their impact on the pendulum performance.

An ion gauge is used to monitor the pressure in the vacuum chamber;
it is only activated occasionally, since continuous operation has been shown to induce charge accumulation on the isolated TMs.

Temperature fluctuations in the laboratory can span several \si{\degreeCelsius} in a weekend and cause variations in the pendulum equilibrium angle at the level of a few hundreds of \si{\milli\rad\per\kelvin}.
In addition, they can affect the gains of the electronics and compromise the stability of the calibrations.
To provide a more stable thermal environment, the vacuum chamber and analog electronics are enclosed in a thermal room.
This room is constructed from wood and foam panels coated with aluminum foil, and provides passive isolation to the experiment.
The reduction in temperature excursions is about a factor 10 at \SI{1}{\milli\Hz} and becomes negligible below \SI{0.1}{\milli\Hz}.
PT1000 temperature sensors read out by a National Instrument NI-9226 board are installed in several locations: on each of the EH, to monitor the temperature gradient within each of them, and on the exterior of the vacuum chamber itself.
The temperature stability measured inside the vacuum chamber is below the white sensor noise floor of \SI{1}{\milli\kelvin\prhz} down to about \SI{2e-4}{\Hz}, where it starts to rise approximately as $f^{-2.5}$.

Magnetic field fluctuations can couple into the pendulum torque noise through the residual magnetic moment and susceptibility of the inertial member. A Barrington Instruments Mag690 3-axial magnetometer has been installed below the center of the vacuum chamber to measure these fluctuations.
The magnetometer exhibits a flat instrumental noise spectrum of \SI{600}{\pico\tesla\prhz} down to approximately \SI{0.1}{\Hz}, when the environmental magnetic field fluctuations start to rise above the instrumental background.

The laboratory is also equipped with a \SI{3}{\m} deep pit, constructed separately from the building's foundation. 
Besides providing additional temperature stability, the pit would also isolate the pendulum from vibrations and seismic activity affecting the building.
Although the pendulum has not yet been moved into the pit, it currently rests on the pit's separate concrete block, which provides some mechanical isolation from the rest of the building.
To further reduce vibrations, the roughing pump is installed in an adjacent hallway and connected to the turbo pump via a flexible bellow.

\section{Pendulum Dynamics} \label{sec:dynamics}
In the rotational degree-of-freedom $\phi$ the dynamics of the pendulum is well described by the equation of motion of a harmonic oscillator with internal damping:
\begin{equation}\label{eq:phiEqMotion}
	- I_r \omega^2 \widetilde{\phi}(\omega) = -\Gamma (1+i\delta) \widetilde{\phi}(\omega) + \widetilde{\tau}(\omega),
\end{equation}
where $I_r$ is the moment of inertia about the rotational axis defined by the fiber, $\Gamma (1+i\delta)$ is the dissipative complex rotational stiffness with small, frequency independent loss angle $\delta$,
and $\widetilde{\phi}(\omega)$ and $\widetilde{\tau}(\omega)$ are the Fourier transforms of the pendulum angular coordinate and of the external torque respectively.
\cref{eq:phiEqMotion} can be recast as:
\begin{equation}\label{eq:phiTF}
\widetilde{\phi}(\omega) = \frac{\widetilde{\tau(\omega)}}{\Gamma \left(1-\left(\frac{\omega}{\omega_0}\right)^2 +\frac{i}{Q}\right)}  \equiv H_{\tau,\phi}(\omega) \widetilde{\tau}(\omega),
\end{equation}
where $\omega_0 = \sqrt{\Gamma/I_r}$ is the natural angular resonant frequency of the torsion pendulum, $Q = 1/\delta$ is the mechanical quality factor, and we have defined the transfer function $H_{\tau,\phi}(\omega)$ between external torque and angular deflection.

The rotational stiffness of the fiber is given by:
\begin{equation}
	\Gamma = \frac{ \pi r^4 }{2L} \left( E_r + \frac{ mg }{ \pi r^2 } \right).
\end{equation}
Here, $r$ and $L$ are the radius and the length of the fiber, respectively, $E_r$ is the shear modulus of the fiber material, and $m$ is the mass of the inertial member. For the parameters listed in \cref{tab:params}, the stiffness is dominated by the shear modulus of the fiber material while the second term represents a small correction of order \SI{1}{\%}.

From the point of view of the swing motion, our apparatus is composed of multiple spherical pendulums in series.
However, the description can be substantially simplified by observing that the size and mass of the eddy current damper's disk are negligible, and only impact the dynamics at frequencies well above our measurement band.
We can thus describe the system as a double pendulum with a massless string (whose length $L_t$ is equal to the length of the two suspension fibers and the disk assembly) suspending a rigid body from a point at a height $h$ above its center of mass.
We call \textit{swing} the deflection of the fiber, and \textit{tilt} the rotation of the inertial member around the attachment point with the fiber.
These two degrees of freedom are mixed in two normal modes.
Even if the small ratio between $h$ and $L_t$ in our geometry makes the mixing ratio relatively small, correctly calculating the resonant frequencies, especially for the mode dominated by the \textit{tilt} motion, requires solving the coupled pendulum equation.

The main observable effect of the \textit{swing} motion is that of translating the inertial member in the horizontal plane.
Due to the length of the suspension fiber, the translation is significant even for comparatively small deflections of the fiber, and the associated rotation can thus be neglected.
Furthermore, we are mainly interested in the motion along the sensitive $x$ direction, since the identical motion in the $y$ direction does not couple directly into the $x$ degree-of-freedom of the TMs.
The translational motion in the $x$ direction can be described as that of a harmonic oscillator.
The \textit{tilt} mode induces both a differential vertical motion and a common mode displacement of opposing TMs. However, the value of the parameter $h$ has been chosen to maximize its frequency and move it outside of the measurement band of interest.
The elasticity of the fiber is associated with the \textit{bounce} mode that causes the entire inertial member to oscillate in the vertical direction.

Ideally, none of these degrees of freedom couple into $\phi$.
However, the high amplitude motion at the respective resonance frequency of each of these degrees of freedom is typically also visible in $\phi$ due to for example static misalignments between the test masses and the electrode housings, or imperfect common mode rejection.
During the design process, we used approximate formulas to estimate the resonant frequencies of the different modes, so that the corresponding peaks could be identified in the pendulum spectrum.
\cref{tab:natFreq} summarizes the estimated and measured values for the most interesting pendulum modes.
\begin{table}
	\caption{\label{tab:natFreq}
		Models and calculated and measured natural periods of the most relevant modes of the torsion pendulum. Parameters are listed in \cref{tab:params}}
	\begin{tabular}{llll}
		\hline\noalign{\smallskip}
		Mode 	 	& Model						& $f_0$ (predicted)		& $f_0$ (measured) \\
		\hline\noalign{\smallskip}
		Rotation 	& Rotational harm. oscill.	& \SI{393}{\micro\Hz}	& \SI{347}{\micro\Hz} \\
		Swing 		& Double pendulum			& \SI{434}{\mHz} 		& \SI{434}{\mHz} \\
		Tilt 		& Double pendulum			& \SI{1.26}{\Hz} 		& \SI{1.23}{\Hz} \\
		Bounce 		& Linear harm. oscill.		& \SI{6.46}{\Hz} 		& \SI{6}{\Hz} \\
		\noalign{\smallskip} \hline
	\end{tabular}
\end{table}

\section{Capacitive sensing and electrostatic actuation} \label{sec:electronics}
The two simplified EHs installed in the facility can both capacitively sense the position of, and electrostatically actuate, their respective enclosed TM.
By measuring the displacement $x_1$ and $x_2$ of the two opposing TMs, both the inertial member rotation angle $\phi$ and translation along $x$ can be obtained through different linear combinations:
\begin{equation}\label{eq:xphiLinComb}
	\phi = \frac{x_1 - x_2}{a}, \qquad
	x = \frac{x_1 + x_2}{2},
\end{equation}
where $a$ is the crossbar arm length.

\subsection{Capacitive readout} \label{sec:capsensing}
\begin{figure}
	\includegraphics[width = \columnwidth]{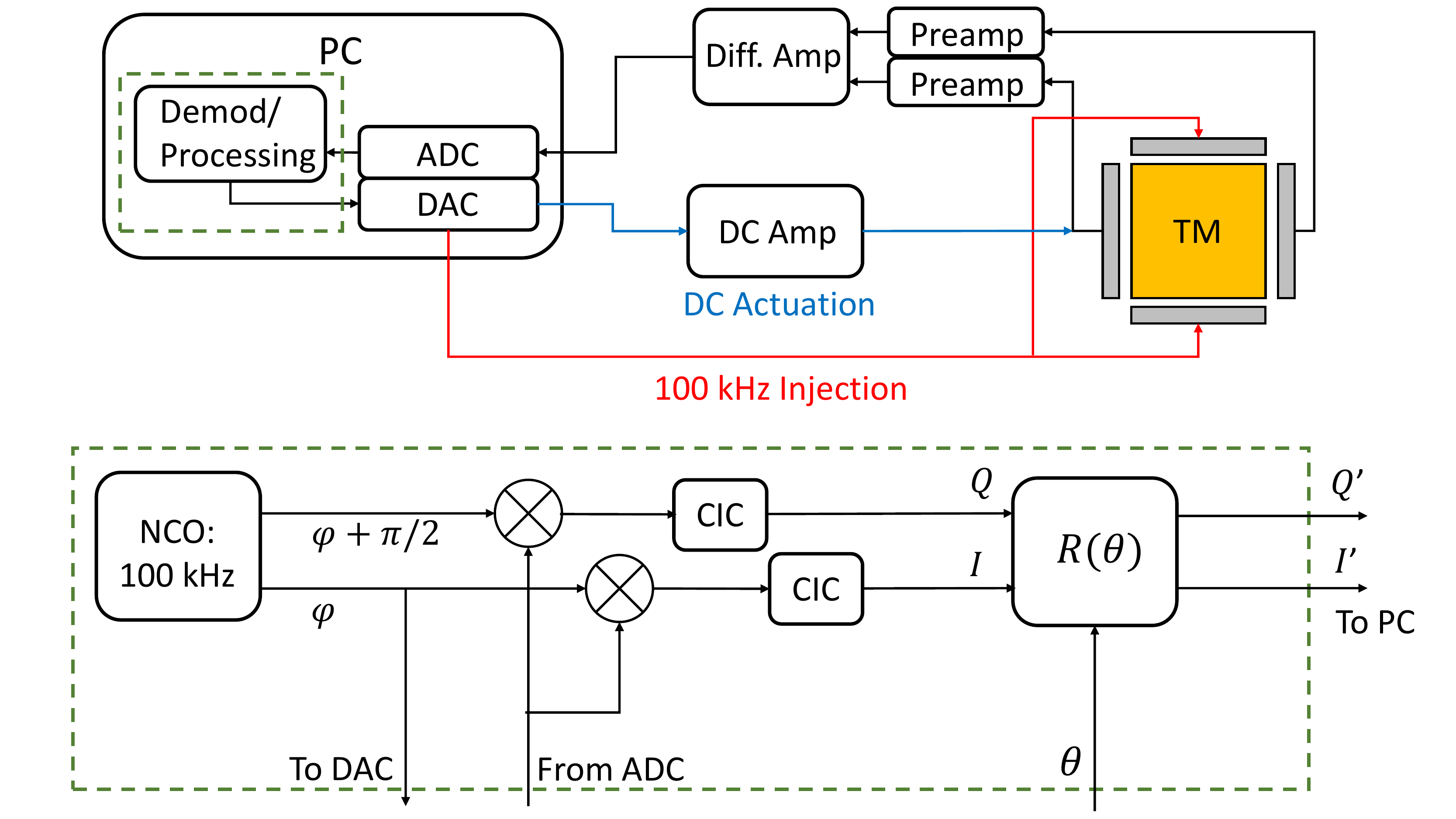}
	\caption{ \label{fig:senseActuate}
		(Color online) Block diagram of the capacitive sensing and electrostatic actuation electronics for a single pair of electrodes (constituting one channel). The TM, surrounded by sensing and actuation electrodes, is represented in the top right corner, together with analog components of the electronics. The upper left corner depicts the digital part, with the digital-to-analog (for generating injection and actuation) and analog-to-digital (for acquiring readout signals) converters and the data processing logic, which is further expanded in the bottom part of the diagram.}
\end{figure}

Using a setup very similar to that employed by LISA Pathfinder\cite{LPF_GRS}, each EH measures the displacement of the enclosed TM by measuring its position-dependent capacitance towards a set of surrounding electrodes.
The TM is polarized with an RF signal using a set of \textit{injection} electrodes.
The readout electronics measures the current flowing to a virtual ground through each of a pair of identical \textit{sensing} electrodes on opposite sides of the TM.
The differential current through the two sensing electrodes is, to first order, proportional to the TM displacement from the nominal center.
\cref{fig:senseActuate} shows a schematic representation of our implementation of this measurement approach.
The injection voltage $V_{inj}$ is a \SI{100}{\kHz} sine wave generated on a National Instruments \mbox{PCIe-8152R} data acquisition (DAQ) board and applied directly to a pair of opposing electrodes on each simplified EH.
In general, the voltage of a TM completely enclosed in an EH is a weighted average of the voltage of the surrounding surfaces:
\begin{equation}\label{eq:VTM}
	V_{TM} =  \frac{\sum_i V_i C_i}{\sum_i C_i} \equiv \frac{\sum_i V_i C_i}{C_{tot}},
\end{equation}
where $V_i$ and $C_i$ are the voltage and capacitance towards the TM, respectively, of the $i$-th surface, and the sum extends over all surrounding surfaces.
For a pair of identical injection electrodes:
\begin{equation}
	V_{TM,inj} = \frac{ 2 C_{inj} }{ C_{tot} } V_{inj} \equiv \alpha V_{inj},
\end{equation}
where $C_{inj}$ is the capacitance between the TM and one electrode.

A pair of sensing electrodes is connected to an analog readout board.
A transimpedance stage keeps each electrode at ground potential (unless a deliberate voltage is applied for actuation purposes), so that the current flowing through each is:
\begin{equation}
	I_\pm = \frac{V_{TM,inj}}{i \omega_{inj} C_{s,\pm} }.
\end{equation}
Here $i$ is the imaginary unit, $\omega_{inj}$ is the injection angular frequency and $C_s$ is the capacitance of a sensing electrode.
The $\pm$ subscript is used to distinguish among the two electrodes.
The currents $I_\pm$ depend on the displacement $x$ of the TM from its centered position inside the EH thorough the capacitance of the electrodes, which have a different (and exactly opposite if they are identical) dependence on $x$.
In particular, using an infinite-parallel-plate approximation:
\begin{equation}\label{eq:ippCapx}
C_{s,\pm}(x)\approx \frac{\epsilon_0 A}{d\mp x},
\end{equation}
where $\epsilon_0$ is the vacuum dielectric constant, $A$ is the area of the electrode and $d$ is the gap between electrode and TM when the latter is centered in the housing. 
The two transimpedance stages output voltages, proportional to $I_\pm$, are fed to a differential amplifier, resulting in a \SI{100}{\kHz} output voltage, amplitude-modulated by the motion of the TM.

We perform demodulation digitally on board the DAQ's field-programmable gate array (FPGA).
The analog signal goes through a single-pole antialiasing filter at \SI{350}{\kHz} before being acquired at about \SI{700}{\kHz}.
It is then mixed with both the \SI{100}{\kHz} digital sinusoid used to generate the injection voltage, and a copy phase-shifted by \SI{90}{\degree}.
Cascaded Integrator-Comb filters low-pass and down-sample the demodulated signals by a factor of 8192, obtaining a final data rate of 85.6 Hz.
The result is finally passed to the host computer for visualization and storage.
By observing the In-phase/Quadrature vector swept out by the pendulum's motion, we can rotate the demodulation phase such that all of the information is in one quadrature to maximize the signal-to-noise ratio.
The capacitive readout is currently limited by the noise in our analog electronics, which is flat in the whole measurement band at a level equivalent to about \sensCap{} per each sensor. This corresponds to a differential capacitance sensitivity of \SI{\sim 7e-18}{\F\prhz} for an injection voltage amplitude of \SI{5}{\V}, corresponding to about \SI{0.75}{\V} at the TM.

\subsection{Actuation} \label{sec:actuation}
The output channels of the DAQ are used to apply actuation signals to the pendulum by varying the voltages of the sensing electrodes.
Applying a voltage difference to two plates of a capacitor results in an attractive force between them.
For a TM enclosed in an EH, one must account for all the internal surfaces and their possibly different potentials.
Doing so, the force along the $x$ direction becomes:
\begin{equation}\label{eq:FxTM}
	F_x =\frac{1}{2}\sum_i \frac{\partial C_i}{\partial x}  \left(V_{TM}-V_i \right)^2,
\end{equation}
where $V_{TM}$ is the TM potential defined by \cref{eq:VTM}. Note that, in calculating the force, the dependence of $V_{TM}$ on $x$ through the $C_i$ (see \cref{eq:VTM}) can be neglected, as its contribution exactly cancels out in the summation.
If we apply a voltage to a single $x$ electrode when the TM is in the electrostatic center:
\begin{equation}\label{eq:FxTMsEl}
F_{x,\pm} \approx \pm \frac{V^2}{2} \frac{C_s(0)}{d} \left(1-\frac{C_s(0)}{C_{tot}}\right),
\end{equation}
where we have used \cref{eq:ippCapx} for the electrode capacitance; the term in parenthesis accounts for the TM polarization due to the actuation voltage.
Analogous expressions can be written for the forces and torques along the other degrees of freedom.

We normally use the \SI{\pm 10}{\V} output of the DAQ without amplification, which according to \cref{eq:FxTMsEl} gives us a maximum force of \SI{\sim 5}{\nano\newton}, or a maximum torque of \SI{\sim 1}{\nano\newton\meter}, per single electrode.
When more force authority is needed, e.g. to stabilize the pendulum when it is out of control, we send the signal through high voltage amplifiers (up to \SI{1}{\kV} maximum output, which amplifies the force by a factor \num{e4}).
The actuation signals are applied to the electrodes through a low-pass filter to reduce transient interference with the readout.

\section{Interferometric Readout} \label{sec:ifo}
\begin{figure}
	\includegraphics[width = \columnwidth]{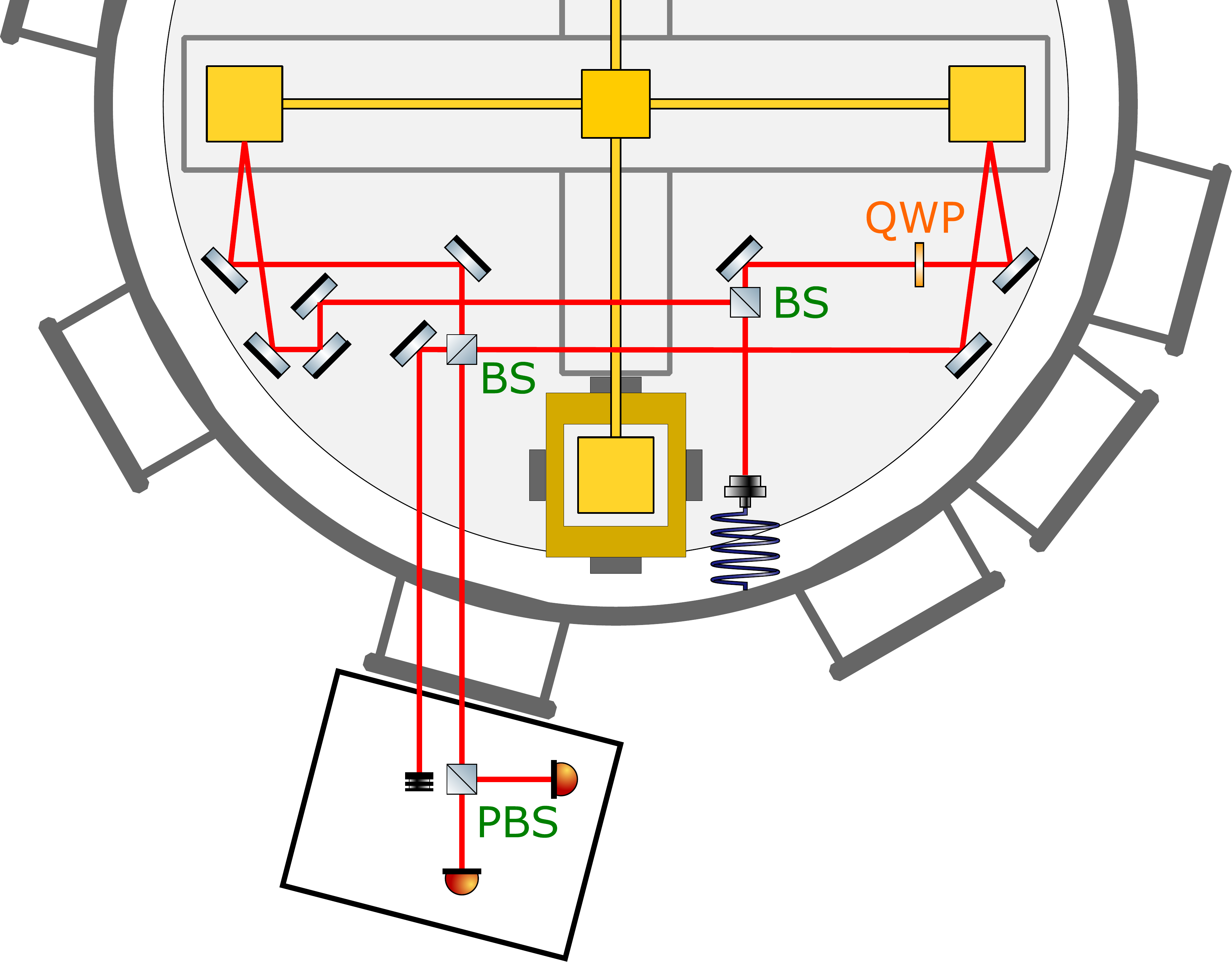}
	\caption{\label{fig:ChamberIFOFancy}
		(Color online) A schematic layout of the readout interferometer. The inertial member and the round platform supporting the sensors are seen from above, together with the electrostatic shields and the outer walls and flanges of the vacuum chamber. One of the EHs in shown in the bottom-center of the picture, while the two TM used for the interferometric measurement are visible on the top, to the left and right. The light enters the vacuum chamber through an optical fiber visible on the right of the EH. The text labels ``QWP'', ``BS'' and ``PBS'' indicate quarter-wave plate, power and polarizing beam splitters, respectively.}
\end{figure}

An alternative readout of the pendulum rotation angle is provided by an interferometer, shown schematically in \cref{fig:ChamberIFOFancy}.
In analogy to \cref{eq:xphiLinComb}, the interferometer is designed to monitor the differential displacement of the two TMs along their $x$ direction while rejecting common mode motion.
The direction of incidence on each TM is chosen to preserve the overlap of the two beams, and thus the contrast, as much as possible upon rotation of the inertial member.
To simplify the optical layout, the entire beam path is on a single horizontal plane at a height slightly below the arms supporting the TMs.

A typical limitation of interferometers is that their sensitivity is greatly diminished around the extremes of a fringe, since the output is 
\begin{equation}
	P_{out} = E_1^2 + E_2^2 + 2 E_1 E_2 \cos(\psi_1 - \psi_2)
\end{equation}
Here, $E_i$ and $\psi_i$ are the amplitude and phase, respectively, of the laser electric field in the $i$th branch of the interferometer.
When $\Delta\psi \equiv \psi_1 - \psi_2 \approx n \pi$, the first derivative of $P_{out}$ with respect to $\Delta\psi$ is zero, and the interferometer looses the ability to track the phase difference.
Consequently, most interferometers are operated at a working point away from $\Delta\psi = n\pi$. 

Such regime is not viable for the torsion pendulum, which we operate without any feedback loop to maintain the lowest possible torque noise.
Even in the quietest state, the differential motion of the TMs is still of the order of several hundred \si{\um} due to both the free natural oscillation and drifts in the equilibrium angle of the fiber.

Following a design developed at the University of Birmingham \cite{PolMuxIFO}, we solve the problem by using polarization-multiplexing to operate two superimposed interferometers whose output is offset by \SI{90}{\degree}.
This allows us to exploit the full sensitivity of each interferometer while still being able to unambiguously follow the motion of the pendulum as it sweeps across many wavelengths of our \SI{1064}{\nm} laser.  
We pass \SI{45}{\degree} polarized light through a power beam splitter, which directs the beam into the two interferometer arms.
In one of the two arms, a $\lambda/4$ wave-plate is inserted with its fast axis aligned with the p-polarization.
This induces a phase delay of \SI{90}{\degree} in that polarization only and in that arm only.
The beams in each arm are reflected off the two opposing TMs which are not enclosed in EHs;
they are then recombined at a power beamsplitter before being sent to a polarizing beamsplitter that separates the two polarizations onto two different photodiodes.
The field in the arm without the $\lambda/4$ wave-plate is 
\begin{equation}
	\vec{E}_1 = E_1 \frac{\hat{p} + \hat{s}}{\sqrt{2}} \mathrm{e}^{i\left(kz - \omega t + \psi_1\right)},
\end{equation}
and the other is 
\begin{equation}
	\vec{E}_2 = E_2 \frac{-i\hat{p} + \hat{s}}{\sqrt{2}} \mathrm{e}^{i\left(kz - \omega t + \psi_2\right)}.
\end{equation}
Here $\hat{p}$ and $\hat{s}$ are the unity vectors of the two polarizations, $i$ is the imaginary unit representing the \SI{90}{\degree} phase shift, $E_i$ is the electric field amplitude in the $i$-th arm, $k = \frac{2 \pi}{\lambda}$ is the wave vector, $z$ is the propagation coordinate and $\omega$ is the angular frequency.

Once the combined output beams have passed through the final polarizing beamsplitter, the photodiode signal for the s-polarization port is
\begin{eqnarray}
	P_{out,s} 	& = & |(\vec{E}_1 + \vec{E}_2) \cdot \hat{s}|^2  \nonumber \\
				& = & \frac{E_1^2}{2} + \frac{E_2^2}{2} + E_1 E_2 \cos\left(\psi_1 - \psi_2\right),
\end{eqnarray}
and for the p-polarization
\begin{eqnarray}
	P_{out,p} 	& = & |(\vec{E}_1 + \vec{E}_2) \cdot \hat{p}|^2  \nonumber \\
				& = & \frac{E_1^2}{2} + \frac{E_2^2}{2} + E_1 E_2 \cos\left(\psi_1 - \psi_2 - \pi/2\right) \nonumber \\
				& = & \frac{E_1^2}{2} + \frac{E_2^2}{2} + E_1 E_2 \sin\left(\psi_1 - \psi_2 \right) .
\end{eqnarray}
In post-processing, we then compute (after subtracting off the DC portion):
\begin{equation}
	\psi_{\mathit{diff},out} = \tan^{-1}\left(\frac{P_{out,p}}{P_{out,s}}\right) = \psi_1 - \psi_2
\end{equation}
In this way, the phase of the output signals can always be unwrapped, because when one quadrature (i.e. one polarization) of the output is at a minimum in sensitivity, the other is at a maximum.
Thus, one can achieve the phase sensitivity of a regular interferometer but with a large dynamic range.
In practice, the displacement of the beams associated with the rotation of the TMs limits the range of our instrument to about \SI{1}{\milli\rad}.

The theoretical sensitivity of the interferometer is set by a number of factors all of the order of \si{\pico\meter\prhz}.
However, current performance appears to be limited by a combination of imperfect common mode rejection of the swing motion, mechanical instability of the components and insufficient mode quality of the beams reflected off of sub-optimally polished TMs.
The readout sensitivity of the differential TM displacement is about \sensIfo{} at \SI{10}{\Hz}, but degrades to about 10 times as much in the \si{\milli\Hz} region, where it becomes comparable with the rotational motion of the pendulum.
A bench-top experiment is being developed to determine to what extent each of the above factors is limiting the performance of the pendulum interferometer, and how the design can be improved.
Results from this experiment, as well as an enhanced TM surface polishing technique, also under development, are expected to improve the interferometer's performance.

\section{Test Mass Charge Measurement and Control} \label{sec:chargecontrol}
Positive and negative charge accumulation on the TM has several detrimental effects on the operation and noise performance of a GRS.
A noisy charging process couples with stray electric fields, which are present inside the GRS due to voltage patches on the internal surfaces, and generates a noisy force on the TM\cite{SpeakePatches}.
Similarly, any temporal variation in the stray electric fields couples with the residual TM charge\cite{MeasDCbias}.
Furthermore, charge accumulation on the TM results in a variation of its potential, and consequently in an electrostatic attraction towards the surrounding surfaces.
For a centered TM, this can be approximated as a negative spring which modifies the dynamics of the system.

In the LISA mission, the maximum allowable TM charge magnitude is \SI{e7}{\elementarycharge} \cite{LISA_CMS}.
Test mass charging is caused by highly energetic particles that penetrate through the spacecraft and either directly, or via secondary electron emission, charge the TM, leading to expected charging rates of \SI{\sim-50}{\elementarycharge\per\second}.
The caging and uncaging process can also leave a residual charge on the TM.

In the UF torsion pendulum, the electrically isolated TM can be left charged by the contact potential when the pendulum is first lifted, by accidental contact with the EH internal surfaces, or by the occasional use of the ion gauge to check the pressure in the vacuum chamber.
A charge management system, capable of measuring and neutralizing the TM charge, is therefore required.

In the LISA GRS, measurement of test mass charge is performed by applying opposite sinusoidal actuation signals to opposing electrodes corresponding to a given degree-of-freedom.
For a centered and discharged TM the attractive forces towards the two opposing electrodes exactly cancel out;
for a charged TM, however, the voltage difference with respect to the two electrodes is no longer symmetric, and the result is  a net force (or torque) at the same frequency as the actuation signal and proportional to the TM charge.

In the torsion pendulum facility, we use the $x$ degree-of-freedom of each TM for charge measurement.
According to \cref{eq:FxTM}, applying a sinusoidal potential $\pm V_{m} \sin (\omega_{m} t)$ to the $+x$ and $-x$ electrodes respectively of a single EH results in a force:
\begin{equation}
	F_{x,m} = 2 \frac{\partial C_s}{\partial x} \frac{q_{TM}}{C_{tot}} V_m \sin (\omega_{m} t) \approx 2 \frac{C_s}{d} \frac{q_{TM}}{C_{tot}} V_m \sin (\omega_{m} t),
\end{equation}
where $q_{TM}$ is the charge on the test mass and we have replaced the value of the $x$ sensing electrode capacitance $C_s$ using \cref{eq:ippCapx}.
Note that, due to the anti-symmetry of the voltages applied to the $\pm x$ electrodes, when the TM is approximately centered in the GRS, its potential is unaffected by the measurement signal.

Moving to the frequency domain and substituting the torque in \cref{eq:phiEqMotion} with $a F_{x,m}$ gives a displacement at $\omega_m$ proportional to the TM charge, with a proportionality constant dependent on the frequency and voltage of the measurement signal: 
\begin{equation}
	\widetilde{\phi}(\omega_m) =2 H_{\tau,\phi}(\omega_m) V_m \frac{C_s}{C_{tot}} \frac{a}{d} q_{TM}
\end{equation}
By demodulating the measured TM displacement at $\omega_m$, one can obtain a dynamic measurement of the TM charge with a bandwidth equal to half the measurement frequency.
Since position readout noise quickly compromises the torque sensitivity of the facility at high frequencies (see \cref{sec:performance}), the choice of $\omega_m$ is a compromise between sensitivity and bandwidth.
We observe that, even if the charge measurement on a single TM results in both a force and a torque on the inertial member, the effect of the former is negligible due to the rigidity of the swing degree-of-freedom compared to the rotational one.

Charge control in the LISA GRS is achieved through UV photoemission\cite{LISA_CMS}:
UV light is directed via two UV fiber feedthroughs toward the TM surface if the goal is to make the TM net charge more positive, or toward the EH surface if the goal is to make it more negative.
For bi-directional charge control using this technique to be successful, the work functions and reflectivities of the test mass and electrode housing surfaces must be roughly the same.
Small levels of surface contamination can greatly alter the work function and therefore the efficiency of the UV photoemission process, hindering the ability to control the TM charge as intended.
In fact not all the light striking a given surface is absorbed; if the reflected light hits a different surface from which it happens to extract electrons with a much greater efficiency and/or a much higher residual kinetic energy, this can result in a net electric charge flow in the unintended direction.

The LISA Pathfinder GRS uses the \SI{254}{\nano\m} UV line of mercury lamps as the light source, but LISA will likely replace the mercury lamps with new deep UV LEDs operating at \SIrange{240}{255}{\nano\m}.
Compared to Hg lamps, UV LEDs are smaller, lighter, consume less power, have a wider spectrum selection, and a higher dynamic range, with at least an order of magnitude improvement in each performance area.
The power output is also very stable, with a lifetime of $\sim30,000$ hours\cite{240LED}.

The faster modulation rate of UV LEDs compared with Hg lamps also means that AC charge control is possible.
In an AC charge control scheme, the UV light is shone on both TM and EH and modulated synchronously with electrode voltages at a frequency that is outside the science band (e.g. the injection electrodes, which already contain a \SI{100}{\kilo\Hz} signal, can be used).
As a result, electrons emitted from any surface will be directed toward the TM or toward the housing depending on the relative phase of the electrode voltage and UV LED modulation.
While the maximum charging rate in each direction for a given UV light power will still depend on the reflectivities and work functions of the surfaces, the AC charge control method should assure that current always flows in the intended direction.
Moreover, the effective rate of charge can be precisely controlled by varying the duty cycle of the UV light emission.

The charge control system of the UF torsion pendulum is similar in design to that of the LISA Pathfinder GRS, but using commercial UV LEDs in a custom assembly:
light from the UV LED is coupled into a fiber optic cable by means of a fused silica bi-convex lens with a focal length of \SI{10}{\mm};
a fused silica ferrule and a borosilicate glass sleeve are used to hold the fiber against the lens and the whole assembly is glued using EPO-TEK 353 ND optical epoxy;
a holder secures the fiber and lens in front of the UV LED.
We also employ a slightly different geometric arrangement of the fibers that direct the UV inside the GRS, with an additional fiber meant to provide a balanced illumination of EH and TM for easier testing of AC discharge.

\begin{figure}
	\includegraphics[width = \columnwidth]{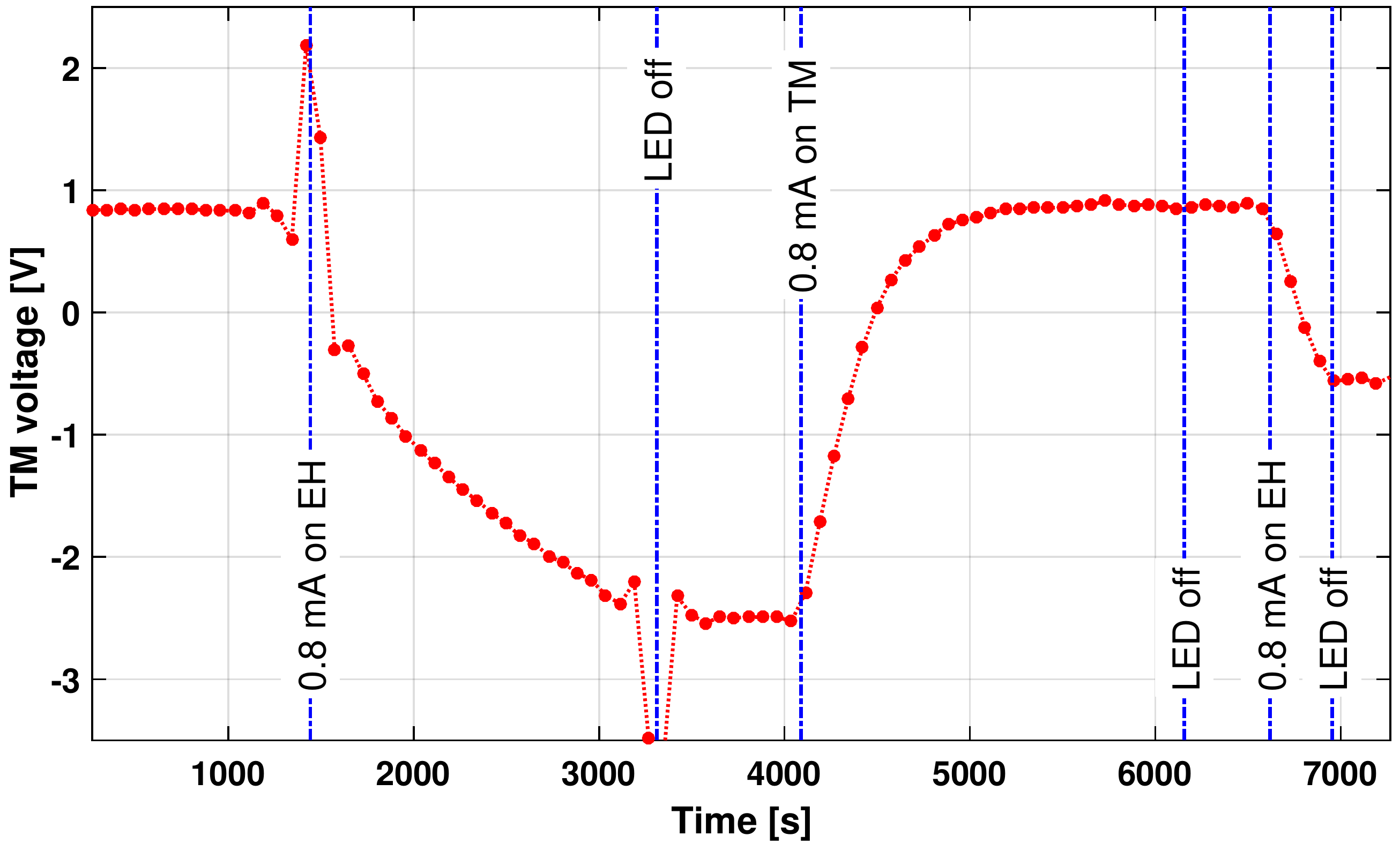}
	\caption{\label{fig:discharge}
		(Color online) Example of charge measurement and control of one of the pendulum TM. The TM potential due to charge, initially measured to be about \SI{0.8}{\V}, is moved to about \SI{-2.4}{\V} over the course of \SI{2000}{\s} using the UV light produced by a UV LED driven by a current of \SI{0.8}{\milli\A} and directed toward the electrode housing. The process is then reversed by illuminating the TM with the same light power. The asymmetry in equilibrium voltages and charge transfer rates in the two cases suggests asymmetry in the illumination pattern and in the reflectivity and work function of the TM and/or EH surfaces. Nevertheless, bipolar charge control is possible. At the time of the measurement, some of the operations indicated by blue vertical lines required someone to physically approach and touch the facility, inducing disturbances in the readout which are visible in the plot as erratic charge values.}
\end{figure}

\cref{fig:discharge} demonstrates effective bipolar charge measurement and control on one of the pendulum TM.
We don't have a way of monitoring the net light output at the end of the fiber, and the \SI{0.8}{\milli\ampere} current used in this example sits right at the knee of the LEDs I/V curve, making it difficult to predict the exact output.
However, despite the current being about a factor 30 below the LEDs' maximum rating, we obtain a charging rate of the order of \SI{5}{\milli\V\per\s}, or \SI{e6}{\elementarycharge\per\s}, which would be sufficient to discharge a LISA TM from its maximum allowed value to zero in about \SI{10}{\s}.
\cref{fig:discharge} also shows that the positive and negative equilibrium voltages are asymmetric around zero.
In principle the equilibrium voltage only depends on the extraction potential for the photoemission process occurring on the surface of the TM and EH, assuming that a few electrons are always extracted from both.
The observed asymmetry thus suggests a difference in the effective work functions of the surfaces, likely caused by variations in finishing and/or contamination.

\section{Instrument performance}\label{sec:performance}
\begin{figure}
	\includegraphics[width = \columnwidth]{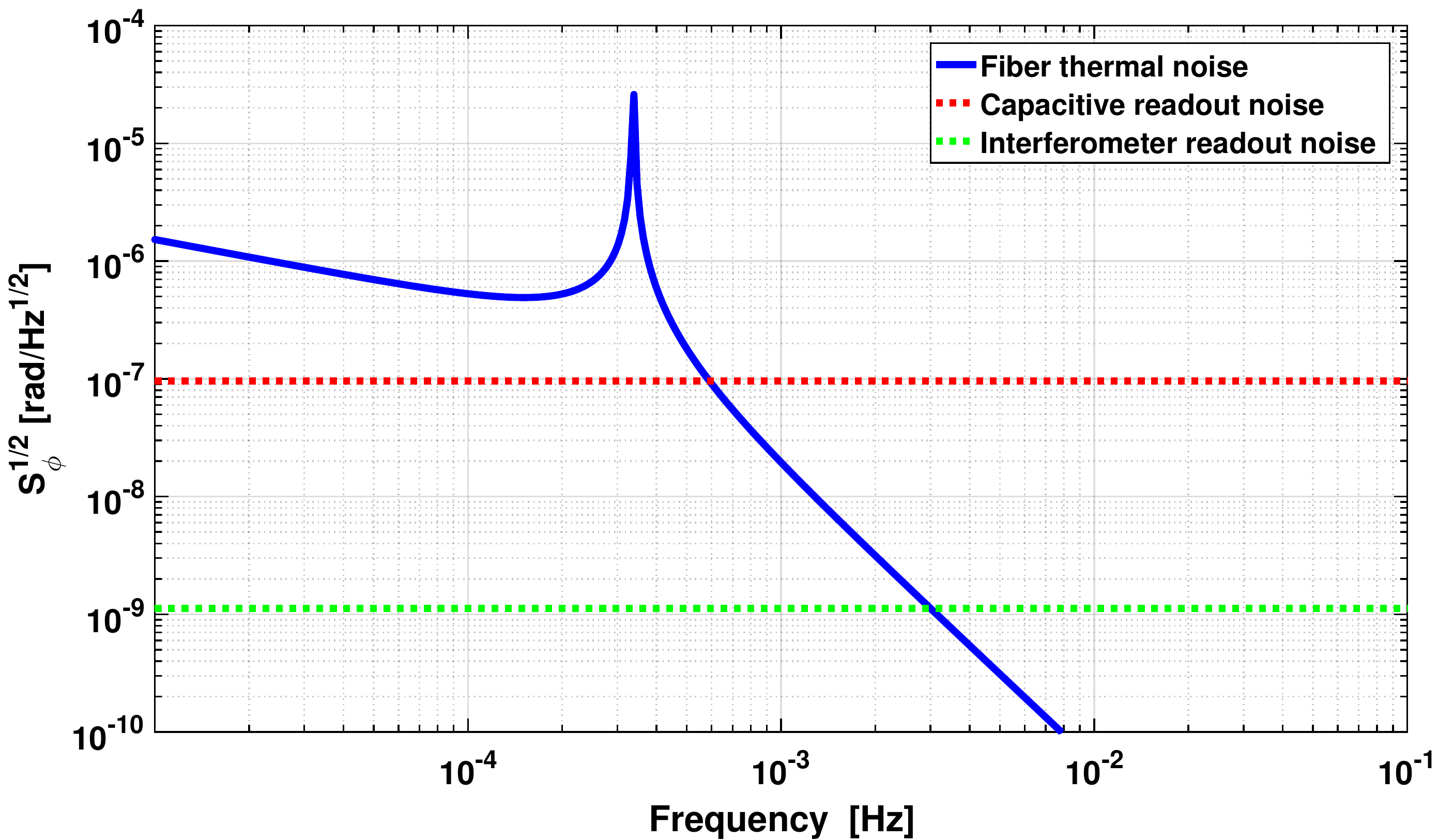}
	\caption{\label{fig:PHINoiseLim}
		(Color online) Displacement noise limit of the facility. Well below resonance, the limit is set by the torque thermal noise in the fiber, converted in angular noise by the flat transfer function of the pendulum; above resonance, both the $1/\sqrt{\omega}$ dependence of the torque noise and the $1/\omega^2$ dependence of the transfer function quickly suppress the angular displacement noise, which rapidly dips below the readout noise, assumed to be white. Given the steep suppression of torque noise at high frequencies, the frequency of the transition from real angular noise to readout noise is not strongly affected by excess torque noise on the pendulum.}
\end{figure}

The sensitivity of the pendulum as a GRS testing instrument is defined by its noise spectrum when a representative GRS is not installed.
The spectrum is limited by both the torque noise acting on the pendulum, usually dominant at low frequencies, and by the readout sensitivity, which dominates at high frequencies.
This is clear when looking at the expected angular noise limit depicted in \cref{fig:PHINoiseLim}.
The fundamental limit to the torque noise is represented by the intrinsic torque thermal noise of the torsion fiber, with power spectral density:
\begin{equation}
S_{N,th}(\omega) = 4 k_B T \frac{\Gamma}{\omega Q},
\end{equation}
where $T$ is the operating temperature, $k_B$ is the Boltzmann constant and $Q$ the quality factor of the fiber introduced in \cref{eq:phiTF}. 
Based on the decay rate of the free oscillation amplitude, we estimated the overall quality factor at resonance to be about 1500, which is approximately a factor 2 below the best quality factors observed with similar tungsten fibers\cite{TrentoSingleTM}.
It is however worth noting that this is just a lower limit to the quality factor of the fiber, as other sources of damping may contribute to the result.
In particular, the damping due to residual gas at our current base pressure is estimated to be not much smaller than the observed one;
indeed, it would contribute half of the observed overall damping at resonance if the pressure inside the simplified EH units, which we cannot directly measure, were only a factor $\sim 2.5$ higher than what we measure in the chamber.
In the plots of this section we nevertheless show a thermal noise limit calculated assuming that the fiber $Q$ is 1500.
While the fiber thermal noise represents a fundamental limit to the performance of the pendulum, many environmental sources may contribute to the overall torque noise, making this lower limit difficult to reach.

The readout noise shows up as an apparent motion of the pendulum, and can be converted to apparent torque noise by dividing by the transfer function defined in \cref{eq:phiTF}.
Above resonance, a readout noise approximately flat in frequency results in a torque noise that increases as $\omega^2$, quickly dominating the spectrum in the upper part of the measurement band.
The uncorrelated noise in the two capacitive sensors translates to an almost white rotational readout noise of about \sensCapPhi{}.
The estimated noise in the interferometric readout of the differential TM displacement corresponds to about \sensIfoPhi{} around \SI{10}{\Hz} and about a factor 10 worse in the \si{\milli\Hz} region.

\begin{figure}
	\includegraphics[width = \columnwidth]{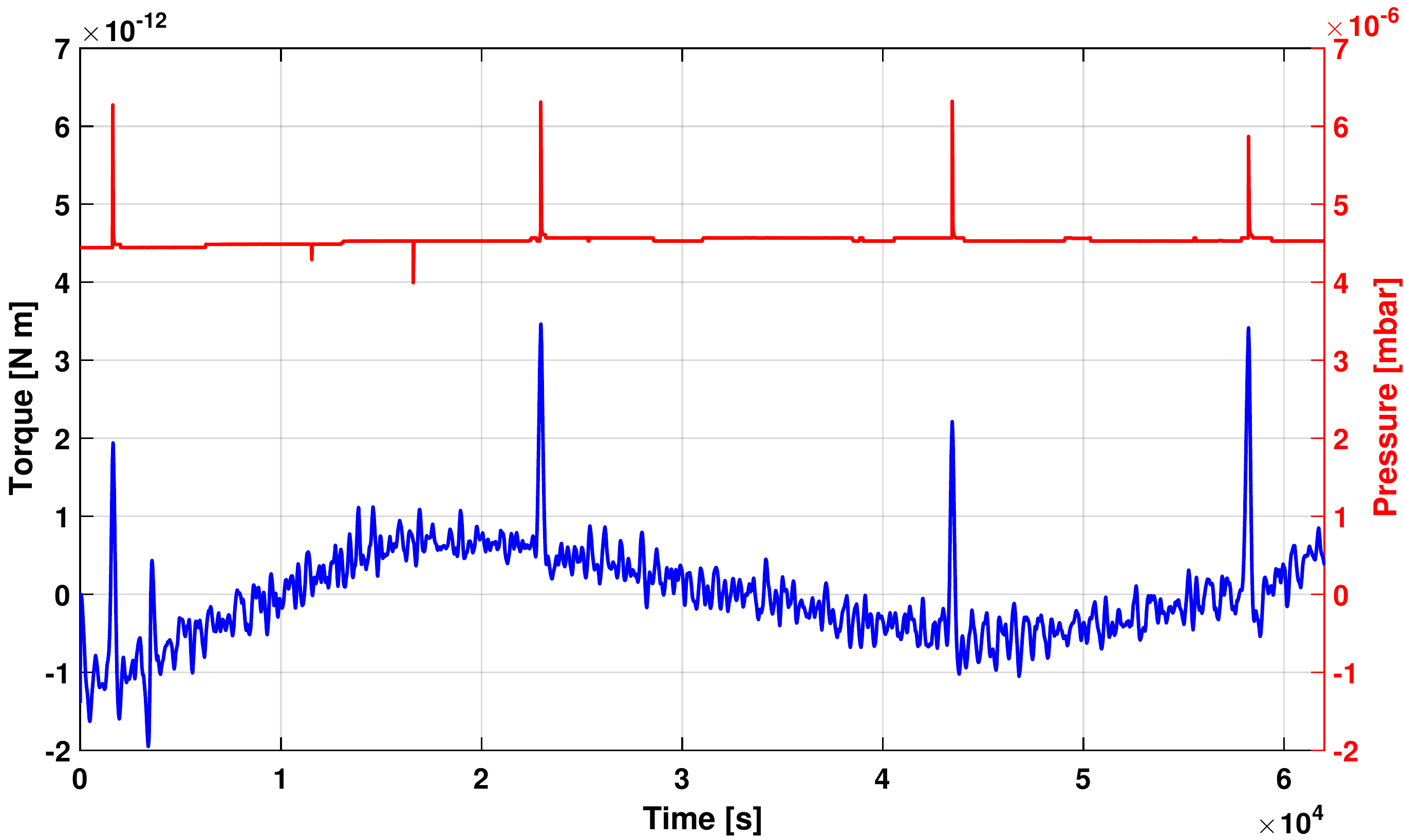}
	\caption{\label{fig:spikes}
		External torque (blue, right axis) and pressure measured by the ion gauge (red, left axis) during a short dedicated run. The correlation between sudden rises in pressure, probably due to a leak, and impulses in the torque time series is evident.
		}
\end{figure}

To evaluate the noise floor of our instrument down to \SI{0.1}{\milli\Hz} or below, we perform dedicated measurements during weekends.
The pendulum natural oscillation is damped to an amplitude of a few tens of \si{\micro\rad} to suppress readout non-linearities, and then left to oscillate freely.
To calculate the external torque from the measured angular time series we use a time domain algorithm\cite{eqmotion} which is part of the LTPDA data analysis package developed for the LISA Pathfinder mission\cite{LTPDA}.
Compared to dividing the angular power spectral density by the torque-to-angle pendulum transfer function, this method has the advantages of being accurate in the presence of transients and much less sensitive to mismatches between real and assumed pendulum parameters, like the quality factor or the free oscillation period.
\begin{figure}
	\includegraphics[width = \columnwidth]{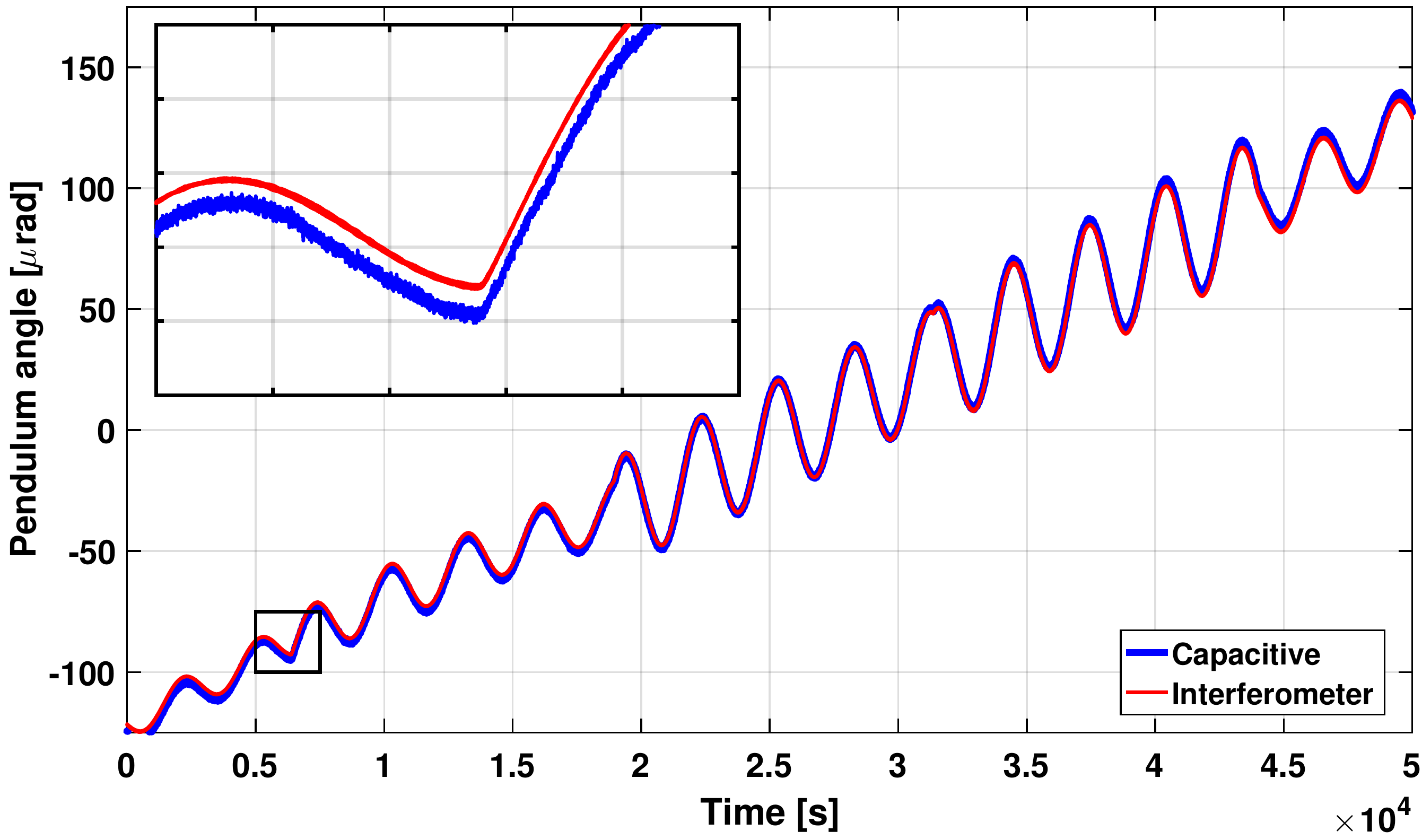}
	\includegraphics[width = \columnwidth]{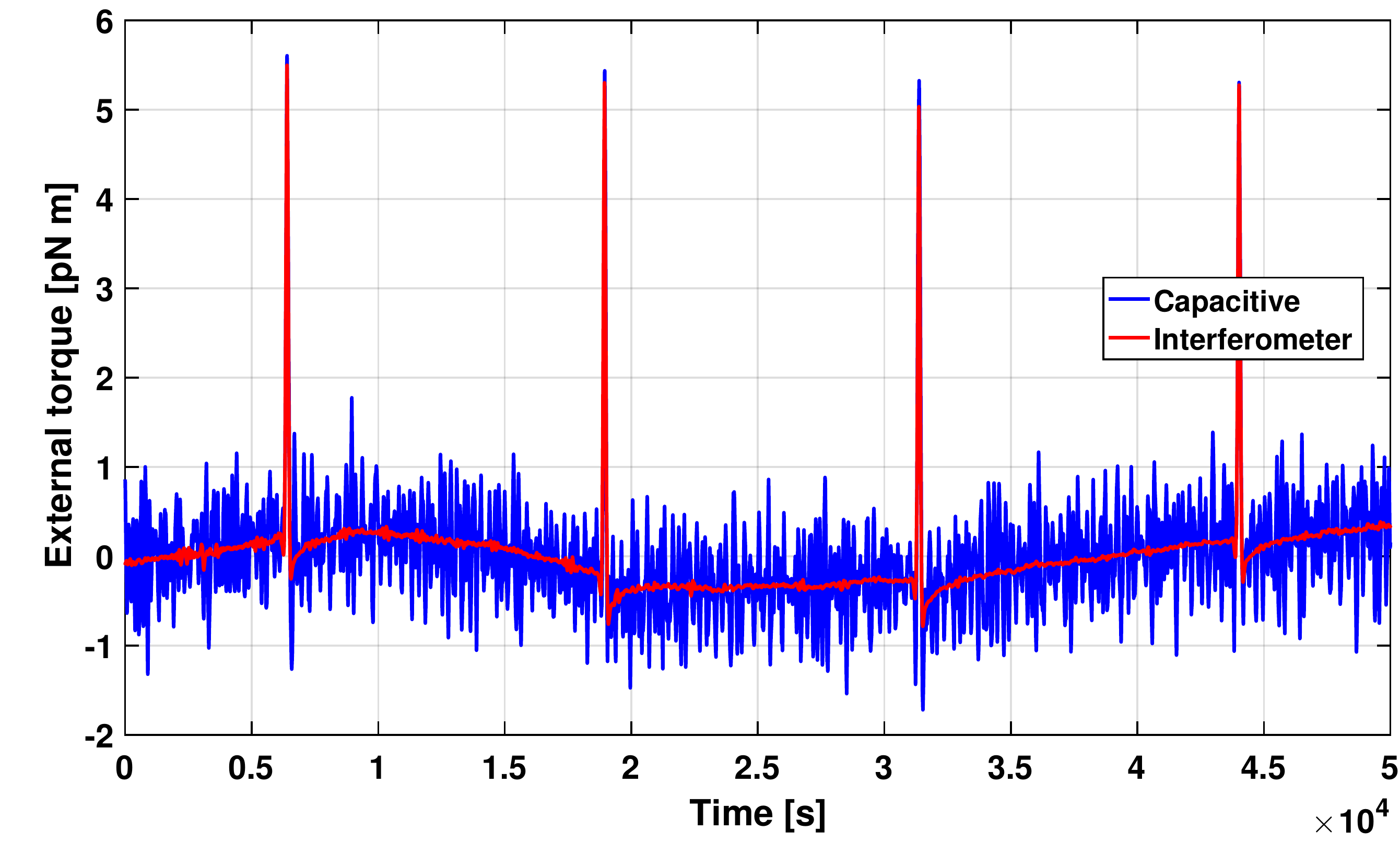}
	\caption{\label{fig:noiseRunTS}
		(Color online) A stretch of 50000 seconds of data during a representative noise run. Sudden changes in the free oscillation phase and amplitude are clearly visible in the angular time series (top). The detail of one of these events is shown in the inset. The slight offset between the capacitive and interferometric traces is due to small drifts in calibration during the entire measurement. Computing the external torque acting on the pendulum (bottom) highlights regular and consistent impulses whose origin has not yet been identified.}
\end{figure}
\cref{fig:noiseRunTS} shows a relatively short stretch of measured pendulum angle $\phi$ from a representative noise run (top plot) together with the calculated external torque, low pass filtered at \SI{5}{\milli\Hz} to suppress the high frequency readout noise (bottom plot).
Almost periodic and identical disturbances are clearly visible in the torque time series.
The period, of about \SI{12500}{\s} in this case, has been observed to vary anywhere between a few thousands to few tens of thousands of seconds, depending on the weekend.
The corresponding angular impulse, however, has a fairly consistent amplitude of the order of \SI{5e-6}{\newton\meter\s} and always points in the same direction.
After a thorough investigation, we have recently determined that the disturbances are due to corresponding bursts in the chamber pressure registered by the ion gauge, as illustrated in \cref{fig:spikes}.
This was not immediately apparent as we usually operate with the pressure gauge turned off to avoid charging of the TM.
Programmatic constraints make it impractical to attempt to address the issue at this time. The facility is about to be partially disassembled to undergo a major upgrade with a consequent downtime of a few months, and we plan to use that occasion to investigate and fix the leak.
In the meanwhile, given the nature of the disturbances, it is reasonable to try to remove their effect when estimating the stochastic torque noise acting on the pendulum, which is the quantity we are interested in characterizing. 
If not accounted for in post processing, these almost-regular impulses give rise to a comb of equal-amplitude harmonics in the power spectral density, practically mimicking a flat noise floor due the lack of resolution.

One simple way of addressing the problem is to limit ourselves to analyzing the data segments in between two adjacent impulses.
While this removes the artificial noise floor, it also severely hinders our ability to investigate noise below a few \si{\milli\Hz}.
\begin{figure}
	\includegraphics[width = \columnwidth]{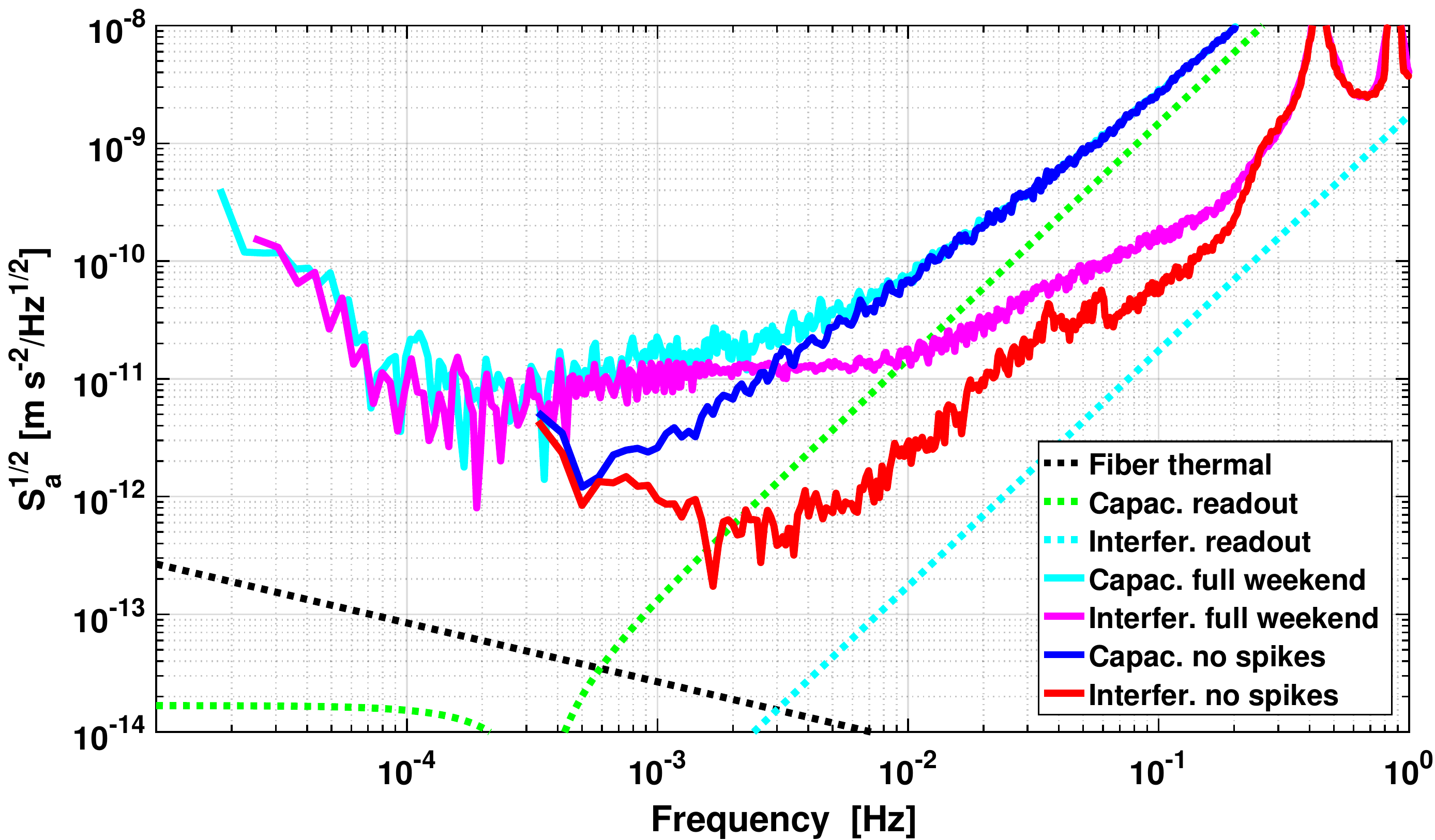}
	\caption{\label{fig:noiseRun}
		(Color online) Fundamental noise limits for the UF torsion pendulum and typical measured torque noise data. Data is presented for both the interferometric and capacitive readout systems. For each, data is analyzed in two different ways to highlight the effect of the disturbance described in the text.
		Note that we produced power spectral density plots using the LTPDA \textit{lpsd} algorithm due to its adaptive averaging capabilities.
		All torque values are converted to equivalent acceleration of a LISA TM by dividing by the pendulum arm length and by the \SI{1.9}{\kg} mass of the TM.}
\end{figure}
\cref{fig:noiseRun} shows the instrumental noise limits discussed above and the external torque power spectral density for a noise run of about 45 hours, from November 19 to November 21, 2016.
Data from both the capacitive sensors and the interferometric readout are used.
For each type of sensor, the data is analyzed both as a single run for the entire weekend, reaching to lower frequencies but showing the artificial floor due to the angular impulses, and in shorter stretches to eliminate their effect.
In the latter case, only segments longer than \SI{12000}{\s} are selected, trimmed to the same length and averaged together.
In all spectra, we have removed the lowermost three frequency points to avoid possible bias due to the spectral estimation window.
We interpret the measured residual torque noise as an upper limit on the force noise acting on a single TM; following a common convention in the LISA community, we express this in terms of equivalent acceleration noise $S_a^\frac{1}{2}$ on a \SI{1.9}{\kg} LISA test mass:
\begin{equation}
S_a^\frac{1}{2} = S_N^\frac{1}{2}\frac{1}{\SI{0.222}{\m}}\frac{1}{ \SI{1.9}{\kg}},
\end{equation}
where $S_N^\frac{1}{2}$ is the torque noise, and in the case of readout noise $S_{\phi,ro}^\frac{1}{2}$:
\begin{equation}
S_{N,ro}^\frac{1}{2} = \frac{S_{\phi,ro}^\frac{1}{2}}{H_{\tau,\phi}(\omega)}.
\end{equation}
To better highlight the current performance of the facility, \cref{fig:bestestimate} only shows our best estimate for each frequency region.
For the lower frequencies we use the spectrum obtained by analyzing the capacitive data for whole weekend at once, but excluding the values above about \SI{0.1}{\milli\Hz}, where the effect of the impulses becomes dominant.
For the higher frequencies, we use the spectrum obtained by analyzing the interferometric readout in the time stretches in between impulses, which limits our lower frequency to about \SI{0.3}{\milli\Hz}.
The region between \SI{0.1}{\milli\Hz} and \SI{0.3}{\milli\Hz} is currently inaccessible, although the expectation is that of a smooth noise spectrum which bridges the two estimates.
\begin{figure}
	\includegraphics[width = \columnwidth]{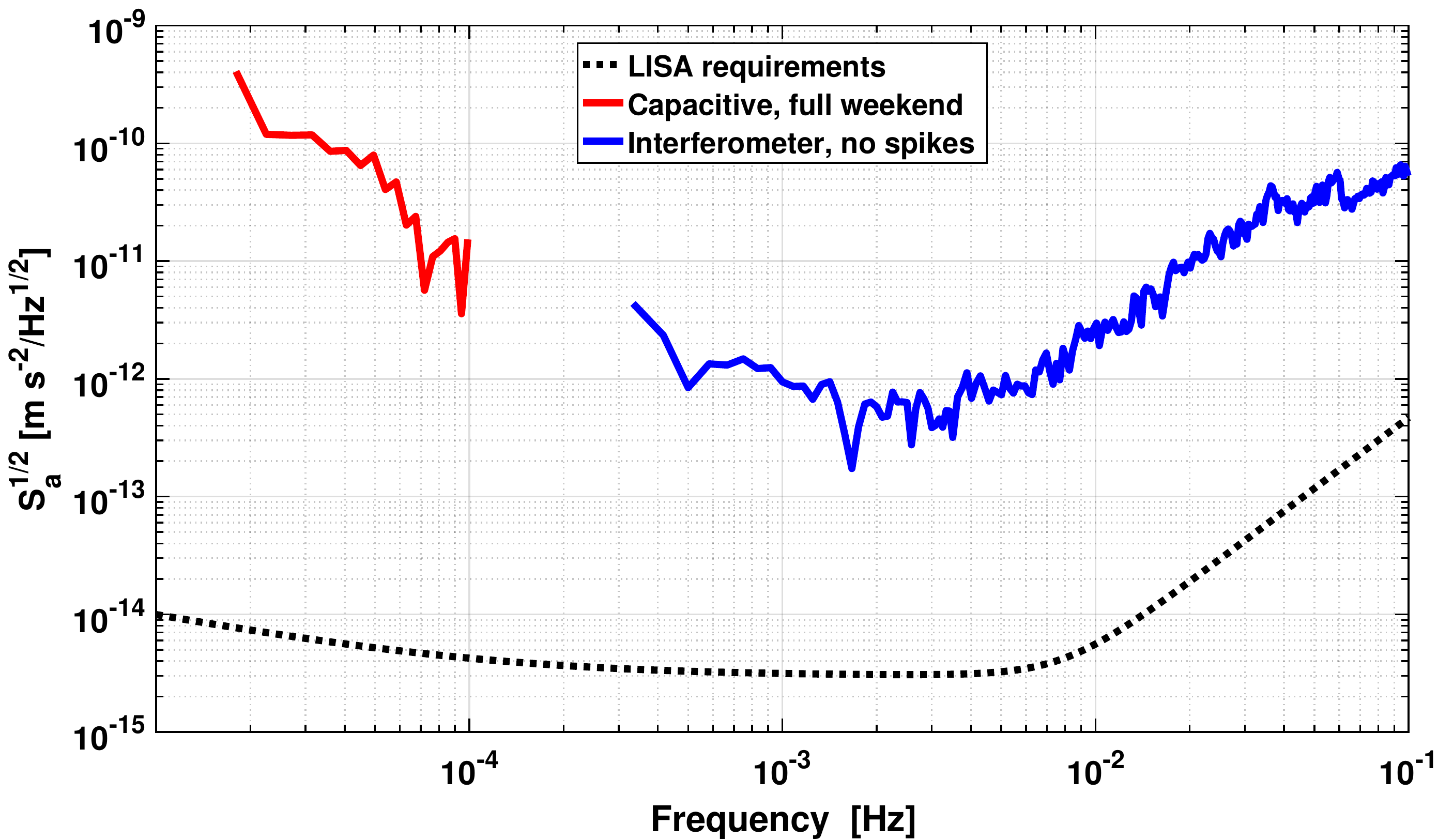}
	\caption{\label{fig:bestestimate}
		(Color online) Best broad-band estimate of the sensitivity of the facility compared to the LISA acceleration requirements. The estimate is obtained by using the interferometric readout spectra shown in \cref{fig:noiseRun}, limited to the frequency regions in which they can be considered representative. For the full weekend spectrum, this corresponds to frequency below about \SI{0.1}{\milli\Hz}, where the pendulum motion is bigger than the floor set by the external angular impulses; for the shorter time stretches, it corresponds to frequencies above about \SI{0.3}{\milli\Hz}, once the lower few points of the spectrum, known to be biased by the spectral estimate, are excluded. The LISA acceleration requirement is also shown for comparison.}
\end{figure}
The impulse-corrected, averaged data shows maximum sensitivities of about \pendBestCapA{} at \pendBestACapF{} and \pendBestIfoA{} at \pendBestIfoF{} using the capacitive and interferometric readout, respectively.
The latter is about a factor \noiseRatio{} above the fiber thermal noise limit, and a factor 100 above the LISA requirements.

\section{Conclusions} \label{sec:conclusions}
We have described a new torsion pendulum facility at the University of Florida designed to test precision inertial sensors for spacecraft with a focus on the gravitational reference sensor for the LISA gravitational wave observatory.
The pendulum features several novel elements, including a laser interferometer readout that is embedded in the pendulum vacuum system and a fiber-coupled UV LED-based charge control system for the inertial sensor test mass.
The pendulum comprises a \SI{1}{\m}-long, \SI{50}{\micro\m}-diameter tungsten fiber that supports a cross-bar with light-weighted test masses mechanically attached to each of its four ends.
The very low torsional spring constant of the fiber and the \SI{22}{\centi\m} arm length of the cross bar results in a near free-fall condition for the test masses in the translational degree-of-freedom orthogonal to both the fiber and the cross bar arm.
Two of the Au-coated aluminum test masses are electrically isolated from the cross-bar and enclosed by electrostatic housings that serve as simplified versions of inertial sensors.
One of these simplified sensors can be replaced by the inertial sensor to be tested.
The sensors capacitively sense the positions of the test masses inside their housings, complementing measurements made by the interferometer, and can exert forces on the test masses using electrostatic actuation.
Each capacitive sensor provides absolute test mass position relative to its electrostatic center to \sensCap{} with a \SI{5}{\V}-amplitude injection signal; the laser interferometer can measure differential displacement of the TMs to better than \sensIfo.

The test mass' potential is measured by applying an oscillating electric field across the test mass using electrodes embedded in the housing and using either the capacitive or interferometric sensors to measure the coherent response of the pendulum.
The amplitude of the pendulum angle at the oscillation frequency is proportional to the test mass charge.
By directing UV light towards the test mass or towards the electrode housing using fiber optic cables connected to UV LEDs, the test mass potential can be increased or decreased respectively, obtaining bipolar discharge.
However, the achievable maximum and minimum values of TM potential are asymmetric around \SI{0}{V} due to differences in the efficiency of the photoemission process occurring on the TM or EH surfaces.
Differences in surface finishing and/or contamination are the mot likely causes of this asymmetry.

The performance of the torsion pendulum measured in terms of the residual torque noise and expressed as equivalent acceleration noise acting on a LISA test mass is \pendBestIfoA{} near \pendBestIfoF{}, which is a factor of roughly \noiseRatio{} above the thermal noise limit of the torsion fiber.
This performance was determined despite systematic torque impulses that were observed in both the capacitive and interferometric readouts.
The impulses occur at nearly regular intervals and have a roughly constant amplitude.
While we have identified the cause of the impulses to be sudden increases in pressure, probably due to a real or virtual leak, we are unable to fix the issue at this time.
Instead, we mitigate their impact on the estimation of the acceleration noise by analyzing the data between impulses when considering effects above \SI{0.3}{\milli\Hz} and by neglecting them when considering effects below \SI{0.1}{\milli\Hz}.
The performance near \SI{0.2}{\milli\Hz} is limited to \SI{10}{\pico\newton\prhz} by these systematic impulses.

In the near future the torsion pendulum facility will be used to evaluate the performance of new technologies and operational modes for precision inertial sensors.
These include new charge control schemes enabled by the relatively high bandwidth of the UV LEDs, which could improve the robustness of the charge control process with respect to variations in the test mass and electrode housing surface properties.
A continuous discharge scheme will also be explored that uses a low level of UV light to keep the test mass potential near \SI{0}{\V} without adversely affecting the acceleration noise of the test mass.

The facility will also be used to evaluate drift mode operation of precision electrostatic accelerometers that can improve their acceleration noise performance without the need for drag-free control and micropropulsion \cite{DMA}.
In a drift mode accelerometer, the electrostatic suspension system is operated with a low duty cycle so that the limiting suspension force noise only acts over brief, known time intervals, which can be neglected in the data analysis.
As a result, drift mode operation has the potential of achieving acceleration noise performance similar to that of drag-free systems although over a more restricted frequency band.

\section*{Acknowledgements}
The authors would like to thank Bill Weber, Rita Dolesi, Antonella Cavalleri, Stefano Vitale, and Daniele Bortoluzzi from the University of Trento, Luciano DiFiore from the University of Naples, and Massimo Bassan from the University of Rome, Sapienza for their valuable technical insight on pendula design, fabrication, instrumentation and operation.
We would also like to acknowledge support from a NASA N.G. Roman Tech Fellowship, grant number NNX15AF26G, NASA grants NNX12AE97G and NNX15AC48G, as well as the Florida Space Grant Consortium.

\bibliography{pendulum}

\end{document}